\documentclass[9pt,twocolumn,twoside]{opticajnl}
\journal{opticajournal}

\setboolean{shortarticle}{true}

\usepackage{upgreek}
\usepackage{bm}
\usepackage{xcolor}

\title{BIC slow light waveguides based on interband coupling}

\author[1]{Sae R. Endo}
\author[1]{Yuta Tanimura}
\author[1]{Takahiro Ito}
\author[2]{Kenta Takata}
\author[3,4]{Takahiro Uemura}
\author[3,4,5]{Masaya Notomi}
\author[6]{Satoshi Iwamoto}
\author[1*]{Yasutomo Ota}

\affil[1]{Department of Applied Physics and Physico-Informatics, Keio University, Yokohama, Kanagawa 223-8522, Japan}
\affil[2]{NTT Research, Inc. Physics \& Informatics Laboratories, 940 Stewart Dr, Sunnyvale, CA 94085, USA}
\affil[3]{Department of Physics, Institute of Science Tokyo, 2-12-1 Ookayama, Meguro, Tokyo 152-8550, Japan}
\affil[4]{NTT Basic Research Laboratories, NTT Corporation, 3-1 Morinosato-Wakamiya, Atsugi, Kanagawa 243-0198, Japan}
\affil[5]{Nanophotonics Center, NTT Corporation, 3-1 Morinosato-Wakamiya, Atsugi, Kanagawa 243-0198, Japan}
\affil[6]{Research Center for Advanced Science and Technology, The University of Tokyo, Meguro, Tokyo 153-8904, Japan}

\affil[*]{ota@appi.keio.ac.jp}

\begin{abstract}
Harnessing bound states in the continuum (BICs) for guiding light in leaky environments has unlocked new possibilities in photonic integrated circuits. BIC confinement enables low-loss waveguiding of leaky TM modes in etchless waveguides based on dielectric wires loaded on plane slabs. We have recently reported BIC slow light waveguides by introducing one-dimensional photonic crystals into such etchless waveguides. However, they were restricted to a high symmetry point (X point), limiting their applicability. In this Letter, we propose and numerically demonstrate BIC slow light waveguides at off-high symmetry points by exploiting Friedrich-Wintgen BICs, arising from the interband coupling of two guided modes sharing a radiation continuum. We identified a systematic approach for tuning the loss minimum position in the momentum space and simultaneously achieved a high group index over $100$ and a low propagation loss of less than $5.0\times10^{-2}$\,dB/cm at an off-high symmetry point. Our findings pave the way for advanced control of light--matter interactions in non-Hermitian photonic systems.

\end{abstract}
\setboolean{displaycopyright}{false} % Do not include copyright or licensing information in submission.

\begin{document}
\maketitle
Bound states in the continuum (BICs), originally proposed in quantum mechanics by von Neumann and Wigner, have been actively studied in optics and photonics. Electromagnetic modes embedded within the continuous spectrum of radiating waves generally experience finite leakage. By contrast, BICs are special modes that lie in the continuum but remain decoupled from radiation, resulting in strong light confinement and high quality factors. Photonic BICs have been realized in various nanostructures, including photonic crystal slabs and metasurfaces\cite{marinicaBoundStatesContinuum2008, plotnikExperimentalObservationOptical2011, hsuObservationTrappedLight2013, hsuBoundStatesContinuum2016, azzamPhotonicBoundStates2021, kangApplicationsBoundStates2023, wangOpticalBoundStates2024}, and have found numerous applications in lasers \cite{kodigalaLasingActionPhotonic2017, hwangUltralowthresholdLaserUsing2021}, filters \cite{foleySymmetryprotectedModeCoupling2014a}, and on-chip photonic devices \cite{bykovBoundStatesContinuum2020}. 

Optical waveguides with BIC confinement have attracted growing interest for enabling low-loss propagation and engineering light--matter interactions in photonic integrated circuits. A typical BIC waveguide is a shallow-etched ridge waveguide supporting transverse magnetic (TM) modes with low effective refractive indices (RIs). The TM modes generally undergo significant loss via coupling to unbound transverse electric (TE) slab modes with high effective RIs. Nevertheless, for certain ridge widths, the TM modes become lossless through BIC confinement induced by decoupling the TM modes from the TE radiation continuum by destructive interference \cite{olinerGuidanceLeakageProperties1981, websterWidthDependenceInherent2007,
koshibaReducedLateralLeakage, nguyenRigorousModelingLateral2009, nguyenLateralLeakageSilicon2020}. Such a BIC waveguide has also been demonstrated in an etchless waveguide based on a low-RI polymer wire loaded on a high-RI plane slab \cite{zouGuidingLightOptical2015, yuPhotonicIntegratedCircuits2019, guRobustBoundStates2023, fengExperimentalObservationDissipatively2023, yuFundamentalsApplicationsPhotonic2023}. This architecture enables tight TM light confinement in etching-challenging materials, offering an uncomplicated scheme to incorporate functional materials into integrated photonics \cite{ yuHybrid2DMaterialPhotonics2019, yuAcoustoopticModulationPhotonic2020, yuHighdimensionalCommunicationEtchless2020, 
yeSecondHarmonicGenerationEtchless2022, zhangUltracompactElectroopticModulator2022, liEfficientSecondHarmonic2022, zhouEfficientOnchipPlatform2025}. 

An approach to further expand the functionality of such BIC waveguide structures involves introducing one-dimensional photonic crystals to the low-RI polymer core loaded on the high-RI material, as recently demonstrated by our group \cite{tanimuraSlowLightWaveguides2025}. Through the periodic modulation along the propagation direction, dispersion curves can be modified to generate modes with low group velocity. Such "slow light" enhances light--matter interactions and contributes to the further miniaturization of photonic devices \cite{babaSlowLightPhotonic2008}. While the proposed structure exhibited BIC slow light modes characterized by low propagation loss and a high group index, these modes were found only at a high symmetry point (X point defined at the zone edge), restricting independent control of propagation loss and group index in momentum space.

In this Letter, we report slow light waveguides based on BICs formed at off-high symmetry points in momentum space. We investigated an etchless waveguide composed of a polymer photonic crystal on a plane slab and utilized the interband coupling between TM guided modes of different orders for creating the off-high symmetric BICs. At optimal geometric parameters, an extremely low-loss mode was observed near the avoided crossing of the coupled guided modes, which we interpret as the emergence of mode-coupling-induced BICs, or Friedrich-Wintgen BICs (FW-BICs). By further tuning the structural parameters, low-loss slow light modes were achieved at off-high symmetry points. Our findings open a new avenue for the application of BICs in integrated photonics.

\begin{figure}[t!]
\centering\includegraphics[width=\columnwidth]{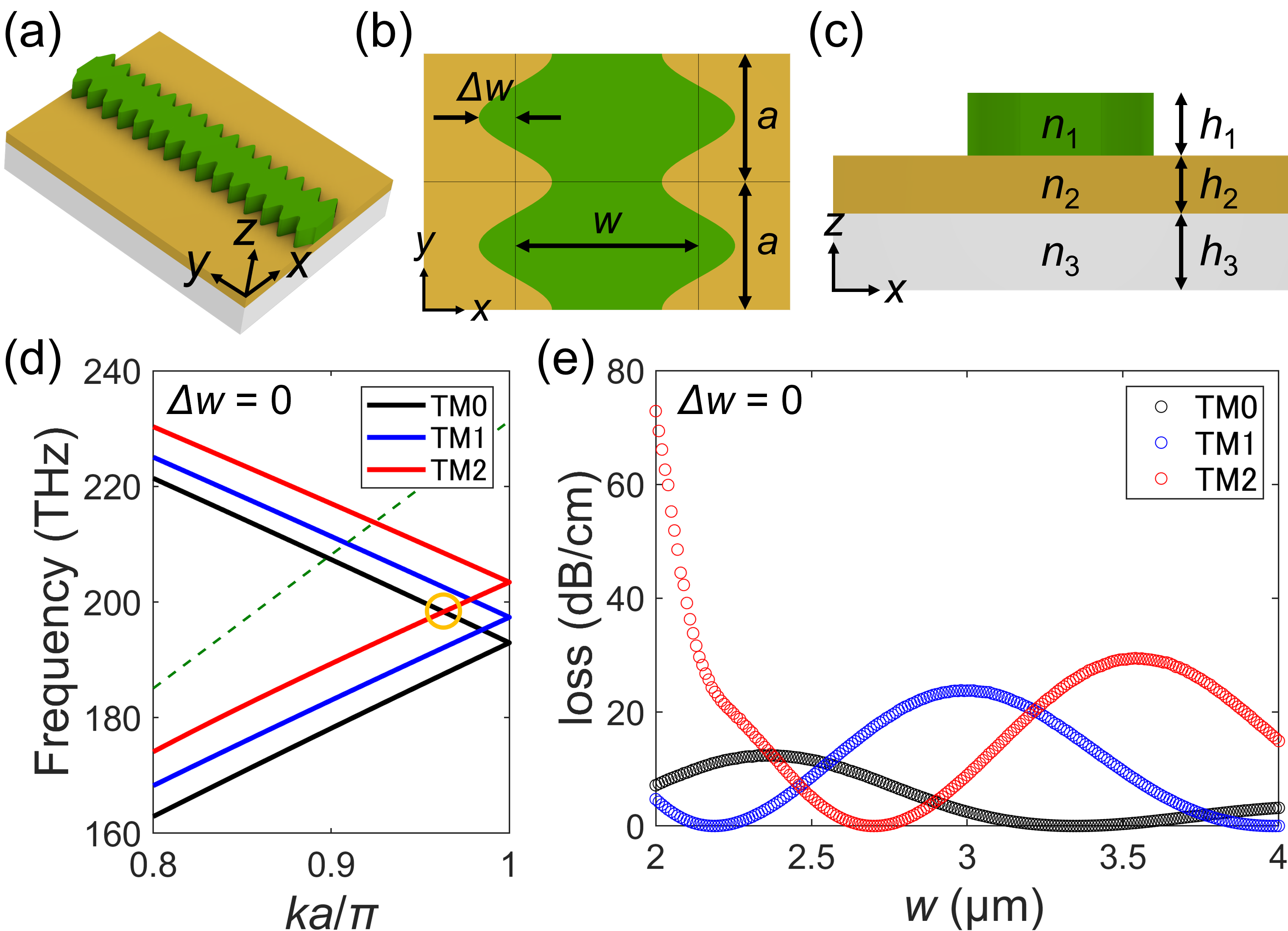}
\caption{(a) Overview, (b) top view, and (c) cross-sectional view of the waveguide structure. (d) Dispersion curves of the unmodulated waveguide ($\Delta w=0$) at $w=2.45$\,$\upmu$m. The green dashed line indicates the glass light line. (e) $w$-dependent propagation loss of the unmodulated waveguide.}
\label{fig:Fig.1}
\end{figure}

Figures 1(a)-(c) illustrate the investigated waveguide structure. A low-RI polymer core ($n_1=1.54$, $h_1=500$\,nm) is loaded on a high-RI dielectric slab ($n_2=2.36$, $h_2=300\,$nm) placed on an SiO$_2$ substrate ($n_3=1.44$, $h_3=2000$\,nm). The width of the low-RI waveguide $w$ is modulated sinusoidally with a depth of $\Delta w$, as depicted in Fig.\,1(b). The period of the modulation is fixed at $a=450$\,nm. Here, TM bound modes, confined beneath the low-RI polymer waveguide via total internal reflection, have lower effective RIs than the TE slab modes and thus experience finite in-plane radiative loss through TE--TM polarization conversion. When the TE--TM coupling is eliminated under BIC conditions, TM modes are laterally well-confined \cite{zouGuidingLightOptical2015, yuPhotonicIntegratedCircuits2019}. The structural and material parameters shown in Fig.\,1(c) can be taken arbitrary to reproduce the physical results presented below, as far as the physical setup does not change significantly. Numerical simulations were performed using COMSOL Multiphysics software based on a three-dimensional finite-element method in the telecommunication wavelength range around $1.5$\,$\upmu$m. Perfectly matched layers were applied along the $x$ and $z$ directions to prevent reflections. Bloch boundary conditions were imposed in the $y$ direction. Photonic band structures, eigenmodes, and propagation loss were computed by the eigenmode solver.

We first investigate the formation of BICs in higher-order TM modes in an unmodulated waveguide ($\Delta w=0$). Fig.\,1(d) shows the dispersion curves of TM$_0$, TM$_1$, and TM$_2$ modes at $w=2.45$\,$\upmu$m, where $k$ is the wave vector in the $y$ direction. Below 210 THz, their frequencies lie under the glass light line indicated in green, confirming the vertical confinement in the slab region. Here, we focus on the intersections of the fictitiously zone-folded lower-order TM modes and the unfolded higher-order TM modes near the X point. They do not interact due to the orthogonality of their wavefunctions in the slab but have finite coupling to the TE slab modes. At certain "magic widths," low-loss BIC TM modes emerge through destructive interference in the TE slab continuum \cite{websterWidthDependenceInherent2007}. Fig.\,1(e) shows the $w$-dependent propagation loss of the unmodulated waveguide at the X point. The oscillatory behavior and the "magic widths" vary with the mode order, implying that the amount of dissipation and their ratio among different order modes can be effectively controlled by adjusting $w$.

\begin{figure}[t!]
\centering\includegraphics[width=\columnwidth]{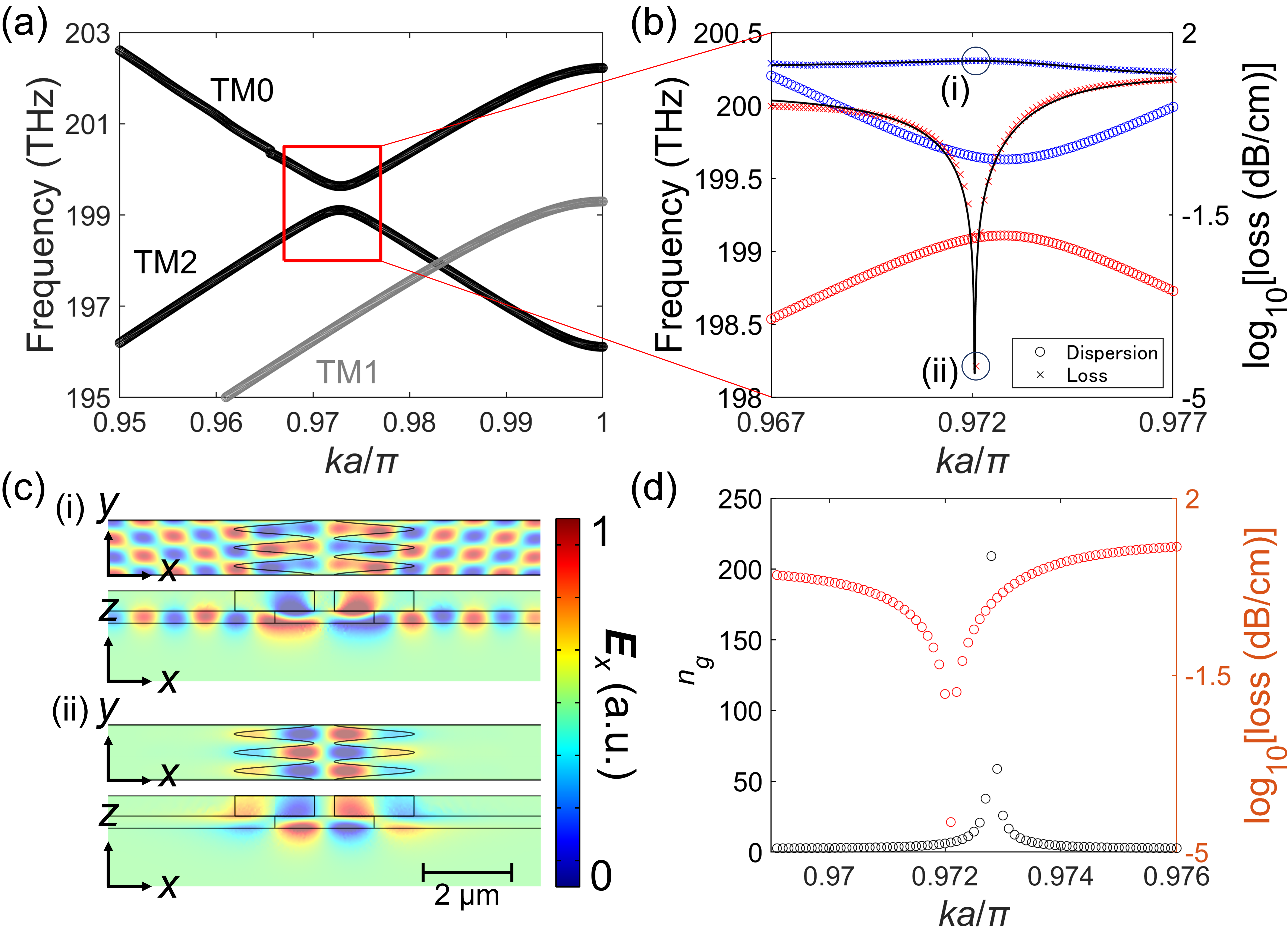}
\caption{(a) Dispersion curves of the modulated waveguide at $w=2.45$\,$\upmu$m and $\Delta w/w=0.40$. (b) Formation of an FW-BIC at an off-high symmetry point with fitting to the theoretical model (black line). (c) Spatial distributions of the $x$-component of the electric field at $ka/\pi=0.9721$. (i) and (ii) correspond to the upper-band lossy mode and lower-band low-loss mode in Fig.\,2(b), respectively. (d) Group index and propagation loss distributions at $w=2.45$\,$\upmu$m and $\Delta w/w=0.40$.}
\label{fig:Fig.2}
\end{figure}

Next, we introduce a periodic modulation along the $y$ direction to enable interactions between TM modes of different orders and induce FW-BICs. We initially set the modulation depth to $\Delta w/w=0.40$ at $w=2.45$\,$\upmu$m. The resulting dispersion curves around the mode intersection (denoted by the orange circle in Fig.\,1(d)) are presented in Fig.\,2(a). The Brillouin-zone-folded TM$_0$ mode and the unfolded TM$_2$ mode interact through the modulation and exhibit avoided crossing behavior at $ka/\pi\approx0.973$. Such near-field coupling occurs only between modes sharing the same parity in the plane normal to the $y$ direction; indeed, a linear crossing is observed between TM$_0$ and TM$_1$ modes at $ka/\pi\approx0.983$. The small gap observed at $ka/\pi\approx0.9655$ on the TM$_0$-like band is attributed to a spurious mode and is not physically meaningful. 

Figure 2(b) shows a magnified view of the dispersions near the avoided crossing with respective loss curves. The upper band corresponds to the lossy mode (i), and the lower band corresponds to the low-loss mode (ii). At $ka/\pi=0.9721$, close to the avoided crossing point, the propagation loss of the lower band drops sharply, reaching a minimum of $3.9\times10^{-5}$\,dB/cm. In contrast, the upper band increases the loss to 29\,dB/cm. The spatial distributions of the $x$-component of the electric field, $\bm{E}_x$, are shown in Fig.\,2(c). While the lossy mode couples to the TE slab mode extending to the edge of the computation domain, the low loss mode is well confined around the low-RI polymer core, indicating that it is a BIC mode where in-plane radiative loss is suppressed by destructive interference.

The emergence of this low-loss BIC mode near the avoided crossing of TM$_0$ and TM$_2$ modes can be interpreted as the formation of an FW-BIC. First proposed by Friedrich and Wintgen \cite{friedrichInterferingResonancesBound1985}, FW-BICs arise from the interference between two coupled resonances sharing a common radiation channel and have been widely studied in photonic systems \cite{rybinHighSupercavityModes2017, azzamFormationBoundStates2018, kikkawaPolarizationbasedBranchSelection2019, kangMergingBoundStates2021}. In our case, the continuum of the TE slab mode serves as the common reservoir for the TM bound modes. In addition to their near-field interaction, radiating waves from the TM$_0$ and TM$_2$ modes into the TE continuum interfere, giving rise to far-field coupling. The interplay between near-field and far-field couplings produces an FW-BIC when the structural parameters are properly tuned.

Such formation of FW-BICs can be described theoretically using a non-Hermitian Hamiltonian. We consider the coupling between TM$_0$ and TM$_2$ modes with eigenfrequencies $\omega_0, \omega_2$ and radiative decay rates $\gamma_0, \gamma_2$ under a shared radiation channel. Then, the non-Hermitian Hamiltonian based on the temporal coupled-mode theory \cite{kangMergingBoundStates2021} can be written as, 
\begin{equation}
    \hat{H}=
    \begin{pmatrix}
        \omega_0 & \kappa \\
        \kappa & \omega_2 \\
    \end{pmatrix}
    -i
    \begin{pmatrix}
        \gamma_0 & \sqrt{\gamma_0\gamma_2}e^{i\psi} \\
        \sqrt{\gamma_0\gamma_2}e^{-i\psi} & \gamma_2 
    \end{pmatrix}, \label{fw-hamil}
\end{equation}
where $\kappa$ and $\sqrt{\gamma_0\gamma_2}\exp{(i\psi)}$ represent the near-field and far-field (via-the-radiation-continuum) couplings, respectively. $\psi$ denotes the relative phase difference of the radiation between the two modes. An FW-BIC is formed when the following conditions are satisfied \cite{kangMergingBoundStates2021}:
\begin{align}
    \kappa(\gamma_0-\gamma_2)=\pm\sqrt{\gamma_0\gamma_2}(\omega_0-\omega_2) \label{fw-biccon},
\end{align}
\begin{align}
    \psi=m\pi\quad(m\in \mathbb{Z}). \label{fw-biccon2}
\end{align}
Then, the eigenvalues of \eqref{fw-hamil} reduce to
\begin{align}
    \omega_+ &= \frac{\omega_0+\omega_2}{2}+\frac{\kappa(\gamma_0+\gamma_2)}{2\sqrt{\gamma_0\gamma_2}(-1)^m}-i(\gamma_0+\gamma_2), \nonumber\\
    \omega_- &= \frac{\omega_0+\omega_2}{2}-\frac{\kappa(\gamma_0+\gamma_2)}{2\sqrt{\gamma_0\gamma_2}(-1)^m}. \label{fw-bic}
\end{align}
Here, one eigenmode ($\omega_+$) becomes more lossy, while the other ($\omega_-$) becomes a lossless BIC mode characterized by a purely real eigenfrequency. For the purpose of demonstrating slow light waveguides, we aim to realize an FW-BIC near the avoided crossing ($\omega_0\sim\omega_2$), which can be achieved when $\gamma_0\sim\gamma_2$ according to \eqref{fw-biccon}. This highlights the significance of controlling $\Delta\gamma =\gamma_0-\gamma_2$. 

We used the theoretical model described in \eqref{fw-bic} to fit the numerically obtained loss curves with $\psi$ fixed at $\pi$ as shown by the black lines in Fig.\,2(b). Here, we assume linear dispersions for the two TM modes when $\kappa = 0$. From the fitting, we extracted near-field coupling strengths and radiative decay rates as $\kappa=0.31$\,THz, $\gamma_0=-0.011$\,THz, and $\gamma_2=-0.0052$\,THz. The simulation results are well described by the non-Hermitian Hamiltonian and satisfy $\gamma_0\sim\gamma_2$, confirming that the observed BIC is indeed an FW-BIC arising from the coupling of TM$_0$ and TM$_2$ modes.

\begin{figure}[htbp]
\centering\includegraphics[width=\columnwidth]{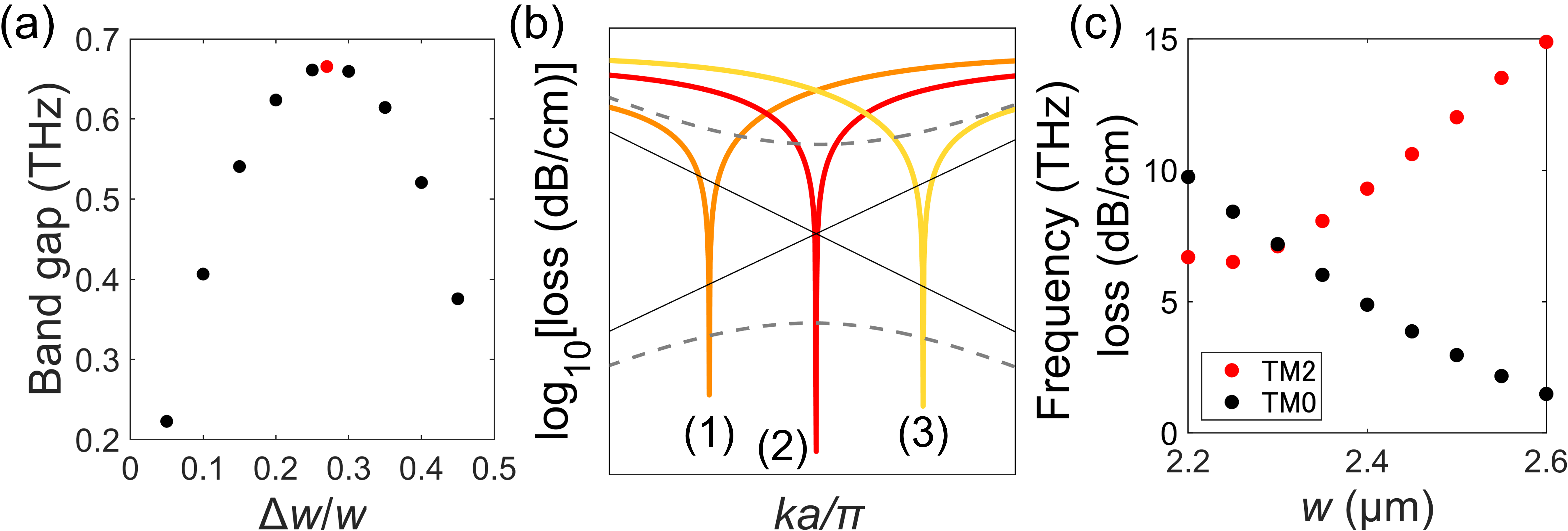}
\caption{(a) $\Delta w$ dependence of the band gap at $w=2.45$\,$\upmu$m. Max band gap is obtained at $\Delta w/w = 0.27$, highlighted in red. (b) Illustration of the FW-BIC position for (1) $\Delta\gamma<0$, (2) $\Delta\gamma=0$, and (3) $\Delta\gamma>0$. The dashed lines indicate the dispersion curves, and the solid lines indicate the lower-band propagation loss. (c) $w$-dependent propagation loss of TM$_0$ and TM$_2$ modes at $\Delta w/w=0.27$ for $ka/\pi=0.975$.}
\label{fig:Fig.3}
\end{figure}

We now examine the slow light behavior in the FW-BIC waveguide. Fig.\,2(d) shows the group index $n_g$ distribution around the avoided crossing with the propagation loss. At the FW-BIC position ($ka/\pi=0.9721$), the group index $n_g$ is $\sim7$, which is significantly higher than that of the host material and demonstrates the behavior of a BIC slow light waveguide at an off-high symmetry point. However, the dominant slow light region is not overlapped with the low-loss region in momentum space. Indeed, the maximum group index $n_g>200$ occurs at $ka/\pi=0.9728$, offset from the FW-BIC position. To resolve the mismatch between high-$n_g$ and low-loss regions, in the following, we will expand the slow light region by enlarging the photonic bandgap and will tune the loss dip to the $n_g$ peak in momentum space. 

As the first step, we maximize the linewidth of the $n_g$ peak. Starting from the FW-BIC condition at $w=$2.45\,$\upmu$m and \mbox{$\Delta w/w=0.40$}, we vary the depth of the sinusoidal modulation $\Delta w$ while keeping $w$ fixed. Fig.\,3(a) shows the $\Delta w$ dependence of the band gap formed by the avoided crossing of the $\mathrm{TM_0}$ and $\mathrm{TM_2}$ modes. The maximum bandgap is obtained at $\Delta w/w=0.27$, highlighted in red, which we assume is linked to the broadest slow light regime. 

Next, we shift the loss dip in momentum space. According to \eqref{fw-biccon}, when two modes have equal loss, i.e. $\Delta\gamma=0$, the loss dip appears at the center of the avoided crossing, where the maximum $n_g$ is available. The movement of the low-loss position with respect to $\Delta\gamma$, calculated from \eqref{fw-bic} with $\psi$ fixed at $\pi$, is demonstrated in Fig.\,3(b). The dashed lines indicate the dispersion curves, and the solid lines show the lower-band propagation loss for different $\Delta\gamma$: (1) $\Delta\gamma<0$, (2) $\Delta\gamma=0$, and (3) $\Delta\gamma>0$. For $\Delta\gamma<0$ ($\Delta\gamma>0$), the FW-BIC forms to the left (right) of the band crossing point. At $\Delta\gamma=0$, the low-loss state is formed precisely at the center of the avoided crossing. In our system, $\Delta\gamma$ can be tuned by adjusting the structural parameters of the low-RI waveguide, as predicted from Fig.\,1(e) with the unmodulated waveguide. With FEM simulations, we confirmed the relative tuning between $\gamma_0$ and $\gamma_2$ for the structure with $\Delta w/w=0.27$ at $ka/\pi=0.975$ as shown in Fig.\,3(c). As $w$ increases, $\gamma_0$ decreases and $\gamma_2$ increases. At around $w=2.30$\,$\upmu$m, they cross, satisfying $\gamma_0=\gamma_2$. This suggests that $\Delta\gamma$ can be systematically controlled by $w$.

Now, we leverage the above behavior to realize the low-loss slow light waveguide with large $n_g$. At $\Delta w/w=0.27$, where we obtained the broadest slow light regime, we tune $w$ to achieve $\Delta\gamma=0$. Fig.\,4(a) shows how the loss dip position shifts in momentum space with $w$. At $w=$2.40\,$\upmu$m, the propagation loss dip lies to the left of the maximum $n_g$ position and moves to the right with increasing $w$, confirming the control of $\Delta\gamma$ by $w$. At $w=$2.47\,$\upmu$m, the loss dip aligns well with the slow light region with the maximum $n_g$ exceeding 100. At the maximum $n_g$, the propagation loss becomes $4.6\times10^{-2}$\,dB/cm, demonstrating slow light waveguiding with negligible dissipation. We note that the depth of the loss dip is shallower than that found in the FW-BIC case discussed in Fig.\,2(d). The deterioration of the loss depth can be attributed to the deviation of $\psi$ from $\pi$ and the formation of a quasi-FW-BIC (qFW-BIC). Indeed, the best fit to the loss curves at $\Delta w/w=0.27$ and $w=$2.47\,$\upmu$m is obtained with $\psi=0.96\pi$. In this case, the radiating TE modes do not form complete destructive interference, leading to finite propagation loss. 

\begin{figure}[t!]
\centering\includegraphics[width=\columnwidth]{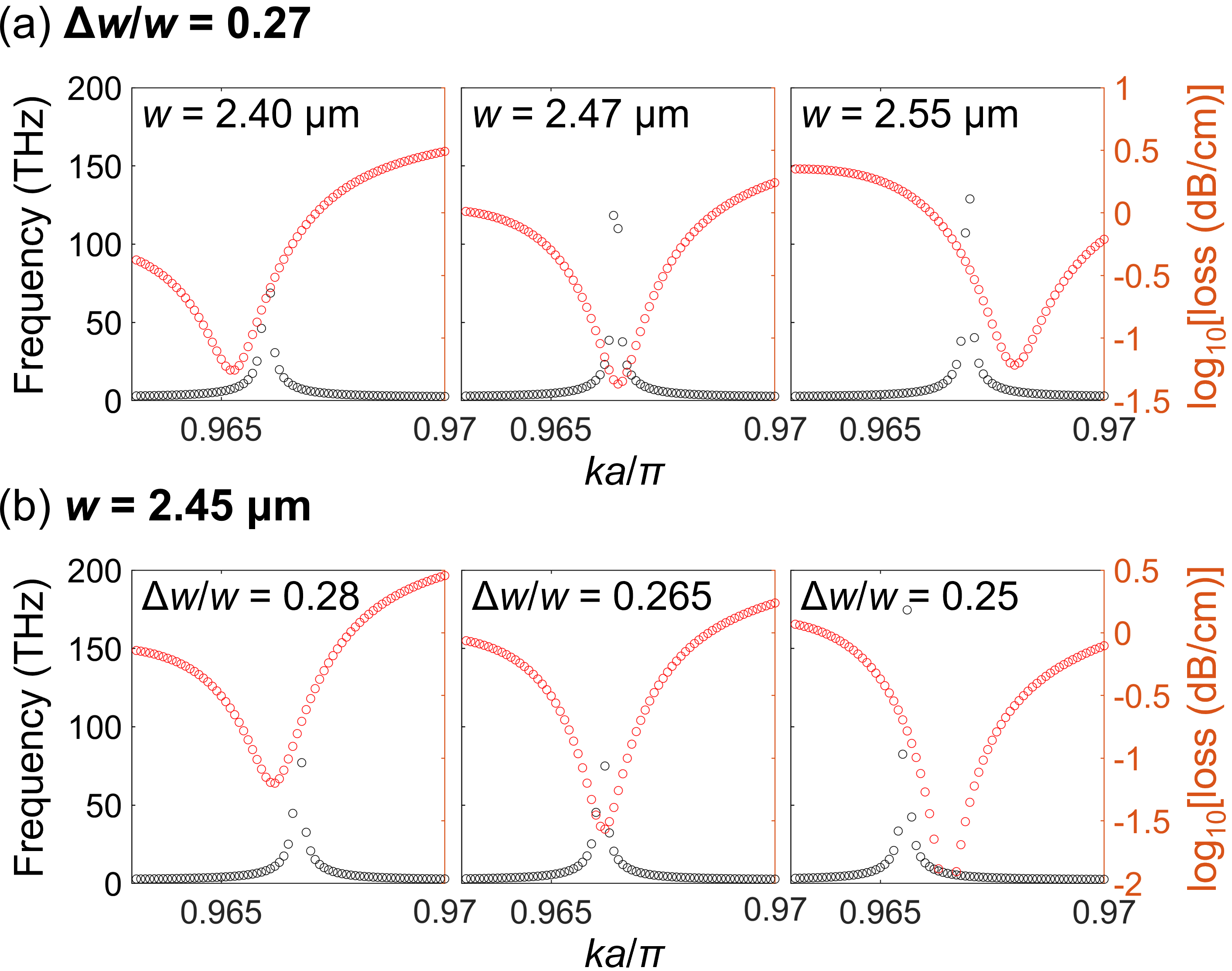}
\caption{(a) qFW-BIC slow light mode achieved by tuning $w$ at $\Delta w/w=0.27$. (b) qFW-BIC slow light mode achieved by tuning $\Delta w$ at $w=2.45$\,$\upmu$m.}
\label{fig:Fig.4}
\end{figure}

Another strategy to induce a sign change in $\Delta\gamma$ and align the FW-BIC with the peak group index is to slightly tune $\Delta w$ at an appropriate $w$, as demonstrated in Fig.\,4(b). Here, we fix the width to the FW-BIC condition $w=$2.45\,$\upmu$m and vary $\Delta w$. At $\Delta w/w=0.28$, the FW-BIC forms on the left side of the max $n_g$ position, similar to the case at $\Delta w/w=0.40$ in Fig.\,2(d). As $\Delta w$ is decreased, the FW-BIC shifts rightward, crossing the max $n_g$ point at $\Delta w/w=0.265$. Here, a qFW-BIC with a propagation loss of $2.7\times10^{-2}$\,dB/cm and $n_g\sim75$ is obtained. Below $\Delta w/w=0.265$, the FW-BIC moves further to the right, away from the slow light regime. Unlike Fig.\,4(a), the loss dip shifts not only horizontally but also vertically as $\Delta w$ is varied. This can be attributed to the high sensitivity of $\psi$ on $\Delta w$ in this regime, affecting the depth of the loss dip (\eqref{fw-biccon2}). 

In conclusion, we demonstrated BIC slow light waveguides at off-high symmetry points. We focused on the interband coupling between TM modes of different orders induced by the periodic modulation of the low-RI polymer core loaded on the high-RI plane slab, which led to the formation of FW-BICs near the avoided crossing. We systematically controlled the positions of FW-BICs in momentum space and achieved high-$n_g$ slow light waveguides with low loss. This approach enables far more flexible control of propagation loss and $n_g$, previously inaccessible with BIC slow light waveguides found at a high symmetry point. Furthermore, the etchless platform studied in this paper is compatible with a broad range of functional dielectric materials. Our work establishes a foundation for exploring advanced functionalities in integrated photonics as well as light-matter interactions with designable non-Hermitian processes.
\\ \\
\noindent\textbf{Funding.}
Fusion Oriented Research for disruptive Science and Technology (JPMJFR213F); Core Research for Evolutional Science and Technology (JPMJCR19T1); Iketani Science and Technology Foundation; Nippon Sheet Glass Foundation for Materials Science and Engineering; Japan Society for the Promotion of Science (22H00298, 22H01994, 22K18989, 25H01400, 25K01697). 

\noindent\textbf{Acknowledgment.}

\noindent\textbf{Disclosures.} 
The authors declare no conflicts of interest. 

\noindent\textbf{Data availability.} 
Data underlying the results presented in this paper are not publicly available at this time but may be obtained from the authors upon reasonable request.

\bibliography{FW-BIC}

\end{document}